\documentclass[%
 aip,
 jmp,%
 amsmath,amssymb,
 reprint,%
]{revtex4-1}

\usepackage{graphicx}
\usepackage{dcolumn}
\usepackage{bm}

\usepackage[cal=boondox]{mathalfa}

\usepackage{mathrsfs}
\usepackage{amsbsy}

\DeclareMathAlphabet\mathbfcal{OMS}{cmsy}{b}{n}

\begin{document}

\preprint{AIP/123-QED}

\title[Quantum Description of FEL Radiation ...]{Quantum Description of Free Electron Laser Radiation and Nonlinear Amplitude Equations}

\author{Stephan I. Tzenov}
\email{stephan@zjlab.ac.cn, \newline {\it Alternative electronic mail}: stephan.tzenov@eli-np.ro} 
\affiliation{Shanghai Synchrotron Radiation Facility (SSRF), 239 Zhangheng Rd, Pudong, Shanghai, China}%
\affiliation{Zhangjiang Laboratory, 99 Haike Rd, Pudong New Area, Shanghai, China}

\author{Zhichu Chen}%
\affiliation{Shanghai Synchrotron Radiation Facility (SSRF), 239 Zhangheng Rd, Pudong, Shanghai, China}%


\date{\today}

\begin{abstract}
A relativistic quantum mechanical model to describe the quantum FEL dynamics has been developed. Neglecting the spin of electrons in the impacting beam, this model is based on the Klein-Gordon equation coupled to the Poisson equation for the space-charge potential and the wave equation for the transverse components of the radiation field. Furthermore, a system of coupled nonlinear envelope equations for the slowly varying amplitudes of the electron beam distribution and the radiation field has been derived. 

The fundamental system of basic equations have been cast into a suitable hydrodynamic formulation. In the framework of the hydrodynamic representation, a new dispersion relation has been derived and analyzed in both the quantum and the quasi-classical regimes, where the space-charge oscillations of the electron beam are taken into account. 

\end{abstract}

\maketitle



\section{\label{sec:intro}Introduction} 

Free electron lasers (FELs) have been invented by John Madey \cite{Madey} in early 1970s, and subsequently have been realized experimentally by his group at Stanford University in the late 1970s \cite{Mad-Group}. A free electron laser is a novel type of light source device. Unlike conventional lasers, where electrons are bound in atomic, molecular or crystalline structures, electrons in a FEL are free in the form of an electron beam accelerated to a suitable energy in a linear accelerator or synchrotron. The conversion of a part of their kinetic energy into coherent electromagnetic radiation takes place in a static magnetic field created by a magnetic device called an undulator. 

FEL devices demonstrate great capabilities and potential as tunable, high-power, coherent sources for short wavelength radiation. In the X-ray wavelength range (from a few nanometers down to 1 Angstrom or less), a high-gain FEL operated in the so-called self-amplified spontaneous emission (SASE) mode can generate multi-gigawatt and femtosecond coherent X-ray pulses. These light sources provide both extremely high power, along with superior transverse coherence properties, which exceed about 10 orders of magnitude in terms of improvement in peak brightness over similar characteristics that can be achieved in modern synchrotron radiation sources based on electron storage rings. Their tunable feature is due to the fact that the wavelength of the coherent radiation is determined mainly by the relativistic electron beam energy and the period of the undulator field. This makes FELs suitable and indispensable experimental means of probing both the ultra-small and ultra-fast worlds.

An important insight gained already in the early days of the FELs development and improvement, especially in the articles by Colson \cite{Colson} and Hopf et al. \cite{Hopf}, was that the FEL dynamics can be completely understood in terms of classical physics. One of the most appropriate methods, in the classical physics framework, to study the dynamics of electrons moving in the undulator field is the kinetic model based on the coupled Vlasov-Maxwell equations. This approach based on the plasma physics description of the underlying processes has been pursued by Davidson and coworkers \cite{Davidson,Davidson_Uhm}. A relativistic cold-fluid dynamics model of an electron beam with uniform cross section propagating axially through a constant-amplitude helical wiggler magnetic field has been developed as well \cite{Davidson_John,Sen_John}. An exact relativistic hydrodynamic closure describing the radiation processes in FELs has been derived by one of us \cite{TzenovMarinov}, and its linear stability has been studied as well.

However, recent years have witnessed a rising interest in a novel regime of FEL operation, where quantum effects become crucial -- the so-called quantum regime or Quantum FEL. In particular, for example, the authors of Refs. \citenum{Bonifacio,Chen} consider the equations of motion of the beam electrons in Heisenberg picture, while in Refs. \citenum{Bonifacio1,Avetissian,Avetissian1} formalism of second quantization is discussed. In addition, a quantum fluid model is developed in Refs. \citenum{Serbeto,Silva} and in Refs. \citenum{Shukla1,Shukla2} the authors use the Klein–Gordon equation coupled to classical electromagnetic fields to derive a dispersion relation. These last two articles are closest in spirit to the exposition presented in what follows. 

Starting from the Hamiltonian governing the motion of a relativistic electron in the static magnetic field of an undulator, and assuming that its dynamics depends on the longitudinal coordinate and the time only, we perform a canonical quantization and arrive at a coupled Klein-Gordon-Maxwell system of equations in one spatial dimension and time. Based on the Renormalization Group (RG) reduction method, we perform in Section \ref{sec:reduction} a coarse-grained transformation of the Klein-Gordon-Maxwell system to a set of nonlinear envelope equations for the slowly varying amplitudes of the electron beam distribution in configuration space and the transverse radiation field. In Section \ref{sec:linearstab}, we first provide a hydrodynamic representation of the Klein-Gordon equation and then perform the linear stability analysis in the proposed fluid dynamic framework. A new dispersion relation has been derived and analyzed in both the quantum and the quasi-classical regimes, where the space-charge oscillations of the electron beam are taken into account. The numerical solution of the dispersion equation thus derived is presented in Section \ref{subsubsec:lindisper} and certain interesting and practically useful properties of the imaginary part of its roots are discussed. Finally, in Section \ref{sec:concluding}, we draw some conclusions. 

\section{\label{sec:basic}Theoretical Model and Basic Equations}

Closely following our previous article \cite{TzenovMarinov}, we start with the description of a relativistic electron beam propagating in the longitudinal $s$ direction through a helical wiggler magnetic field described by
\begin{equation}
{\bf A}_w = - {\frac {B_0} {k_0}} {\left( {\bf e}_x \cos k_0 s + {\bf e}_y \sin k_0 s \right)}, \label{Undulator}
\end{equation}
where $B_0 = {\rm const}$ is the field amplitude, $\lambda_0 = 2 \pi / k_0$ is the wavelength, and ${\bf e}_x$ and ${\bf e}_y$ are unit Cartesian vectors in the plane perpendicular to the propagation direction. Continuing the analogy further, we assume that spatial variations are one-dimensional, so that the electron beam dynamical variables and the radiation field evolve only in the longitudinal direction ${\left( \partial_s = \partial / \partial s \neq 0 \right)}$. This implies that the canonical momenta ${\bf p}_{\perp}$, transverse to the beam propagation direction are integrals of motion. Thus, the Hamiltonian of an electron moving in the wiggler field, as well as in the self-consistent electromagnetic field can be written as
\begin{equation}
\gamma_c {\left( s, p_s; \tau \right)} = {\sqrt{1 + p_s^2 + a_x^2 + a_y^2}} - \varphi - \partial_{\tau} \int {\rm d} s a_s {\left( s; \tau\right)}, \label{HamilA}
\end{equation}
For the sake of convenience dimensionless variables according to the relations 
\begin{equation}
\tau = c t, \qquad \qquad {\bf v} = {\frac {\bf V} {c}}, \label{Variables1}
\end{equation}
\begin{equation} 
{\bf p} = {\frac {\bf P} {m_e c}}, \qquad {\bf a} = {\frac {e \bf A} {m_e c}}, \qquad \varphi = {\frac {e \Phi} {m_e c^2}}, \label{Variables2}
\end{equation}
have been introduced. Here $e$ is the electron charge, $m_e$ is the electron rest mass, $c$ is the velocity of light in vacuo, and ${\bf P}$ and ${\bf V}$ are the actual canonical momentum and the actual particle velocity, respectively. In addition, $\Phi$ is the scalar self-potential and ${\bf A}$ is the total vector potential ${\bf A} = {\bf A}_w +{\bf A}_r$, where ${\bf A}_r$ represents the radiation self-field. 

The scaled dimensionless electromagnetic potentials $\varphi$ and ${\bf a}$ satisfy the equations \cite{TzenovMarinov} 
\begin{equation}
{\boldsymbol{\Box}} a_s = {\frac {\mu_0 e^2} {m_e}} n v_s, \qquad \qquad {\boldsymbol{\Box}} \varphi = {\frac {\mu_0 e^2} {m_e}} n, \label{WaveEqu1}
\end{equation}
\begin{equation}
{\boldsymbol{\Box}} {\cal A} + 2 i k_0 \partial_s {\cal A} - {\left( k_0^2 + {\frac {\mu_0 e^2 n} {m_e \Gamma}} \right)} {\cal A} = {\frac {\omega_c k_0} {c}}. \label{WaveEqu2}
\end{equation}
Here ${\boldsymbol{\Box}} = \partial_s^2 - \partial_{\tau}^2$ denotes the well-known d'Alembert operator. Moreover, 
\begin{equation}
{\cal A} = {\left( a_x + i a_y \right)} e^{-i k_0 s} = {\cal A}_x + i {\cal A}_y, \qquad \quad \omega_c = {\frac {e B_0} {m_e}}, \label{TranVecPot}
\end{equation}
where $\omega_c$ the electron cyclotron frequency associated with the amplitude of the wiggler field. In the above expressions, $\mu_0$ is the magnetic susceptibility, $n$ is the number density, $v_s$ is the longitudinal component of the macroscopic flow velocity of the electron beam, while the quantity $\Gamma$ can be written as \cite{TzenovMarinov} 
\begin{equation}
\Gamma = {\sqrt{\frac {1 + {\left| {\cal A} \right|}^2} {1 - v_s^2}}}. \label{Gamma}
\end{equation}
It is worthwhile to mention that we consider here the simplest case of a cold electron beam \cite{TzenovMarinov}, which is the most common case in practice.

The final step in our preparatory consideration is the canonical quantization of the relativistic Hamiltonian (\ref{HamilA}) by following the general operator substitution rule 
\begin{equation}
\gamma_c \longrightarrow i {\widetilde{\hbar}} \partial_{\tau} \qquad \quad p_s \longrightarrow - i {\widetilde{\hbar}} \partial_s \qquad \quad {\widetilde{\hbar}} = {\frac {\hbar} {m_e c}}. \label{CanonQuantiz}
\end{equation}
Neglecting the spin of the electrons comprising the impacting beam, and omitting (in what follows) the "tilde"-sign over the re-scaled Planck's constant ${\widetilde{\hbar}}$, we can write the Klein-Gordon equation 
\begin{equation}
{\left( i \hbar \partial_{\tau} + V \right)}^2 \Psi + \hbar^2 \partial_s^2 \Psi - {\left( 1 + {\left| {\cal A} \right|}^2 \right)} \Psi = 0. \label{KleinGord}
\end{equation}
for the electron wave function $\Psi$. We have introduced a new generalized scalar potential $V$ according to the expression 
\begin{equation}
V {\left( s; \tau \right)} = \varphi {\left( s; \tau \right)} + \partial_{\tau} \int {\rm d} s a_s {\left( s; \tau\right)}. \label{NewScalarPot}
\end{equation}
Using the Lorentz gauge 
\begin{equation}
\partial_{\tau} \varphi + \partial_s a_s = 0, \label{LorentzGauge}
\end{equation}
together with Eqs. (\ref{WaveEqu1}) it is straightforward to verify that 
\begin{equation}
\partial_s^2 V = {\boldsymbol{\Box}} \varphi = {\frac {\mu_0 e^2} {m_e}} n. \label{LaplaceEq}
\end{equation}
Note that the longitudinal electric force acting on the electrons in the impacting beam is ${\cal F} = \partial_s V$. In this sense the generalized potential $V$ can be regarded as an electrostatic potential describing the space charge.

Rearranging terms in Eq. (\ref{KleinGord}), we can rewrite it as follows 
\begin{equation}
\hbar^2 {\boldsymbol{\Box}} \Psi + 2 i \hbar V \partial_{\tau} \Psi - {\left( 1 + {\left| {\cal A} \right|}^2 - V^2 - i \hbar \partial_{\tau} V \right)} \Psi = 0. \label{KleinGordon}
\end{equation}
Standard manipulation of the Klein-Gordon equation (\ref{KleinGordon}) transforms it into the continuity equation 
\begin{equation}
\partial_{\tau} n + \partial_s {\left( n v_s \right)} = 0, \label{Continuity}
\end{equation}
where 
\begin{eqnarray}
n = {\frac {i \hbar} {2}} {\left( \Psi \partial_{\tau} \Psi^{\ast} - \Psi^{\ast} \partial_{\tau} \Psi \right)} - V {\left| \Psi \right|}^2, \label{Density}
\\
j_s = n v_s = {\frac {i \hbar} {2}} {\left( \Psi^{\ast} \partial_s \Psi - \Psi \partial_s \Psi^{\ast} \right)}, \label{CurVel}
\end{eqnarray}
are the electron current density and $v_s$ is the longitudinal component of the electron current velocity. 

Note that as far as notation is concerned, here we have simply adopted $n$ to denote the quantity $n \Gamma$ of Ref. \citenum{TzenovMarinov}. The set of equations (\ref{WaveEqu2}), (\ref{LaplaceEq}) and (\ref{KleinGordon}) comprises the basic system of equations, which represents the starting point of our subsequent analysis.  

\section{\label{sec:reduction}Renormalization Group Reduction}

The outlined above basic system of equations (\ref{WaveEqu2}), (\ref{LaplaceEq}) and (\ref{KleinGordon}) possesses an exact stationary solution 
\begin{equation}
\Psi_0 = 0, \qquad n_0 = 0, \qquad V_0 = 0, \qquad {\cal A}_0 = A_0 = - {\frac {\omega_c} {c k_0}}. \label{ExactStatSol}
\end{equation}
Following the standard procedure of the Renormalization Group (RG) reduction method \cite{TzenovBOOK} applied to the system of equations (\ref{WaveEqu2}), (\ref{LaplaceEq}) and (\ref{KleinGordon}), we represent the electron wave function $\Psi$, the complex transverse vector potential ${\cal A}$ and the generalized scalar potential $V$ as a perturbation expansion about the exact stationary solution (\ref{ExactStatSol}). In more explicit form   
\begin{equation} 
\Psi = \sum \limits_{k=1}^{\infty} \epsilon^k \Psi_k, \quad {\cal A} = A_0 + \sum \limits_{k=1}^{\infty} \epsilon^k {\cal A}_k, \quad V = \sum \limits_{k=2}^{\infty} \epsilon^k V_k. \label{DenseExpand}
\end{equation}
Here $\epsilon$ is a formal small parameter, which will be set equal to one at the end of all calculations. The next step consists in expanding the system of Klein-Gordon and radiation field equations (\ref{WaveEqu2}), (\ref{LaplaceEq}) and (\ref{KleinGordon}) in the formal small parameter $\epsilon$. Their perturbation solution will be obtained order by order together with performing a procedure of elimination of secular terms (usually completed at third order), which will yield the sought for amplitude equations for the slowly varying envelopes. 

\subsubsection{\label{subsubsec:firstorder}First Order} 

The linearized Eqs. (\ref{WaveEqu2}) and (\ref{KleinGordon}) can be written as 
\begin{equation}
{\boldsymbol{\Box}} {\cal A}_1 + 2 i k_0 \partial_s {\cal A}_1 - k_0^2 {\cal A}_1 = 0, \label{WaveEquFirOrd}
\end{equation}
\begin{equation}
\hbar^2 {\boldsymbol{\Box}} \Psi_1 - {\left( 1 + A_0^2 \right)} \Psi_1 = 0. \label{KleinGordFirOrd}
\end{equation}
The general solutions to the above first-order equations can be expressed in the form 
\begin{equation}
{\cal A}_1 = A_1 e^{i \phi_1} + A_2 e^{i \phi_2}, \qquad \Psi_1 = B_1 e^{i \psi} + B_2 e^{-i \psi}. \label{SolEquFirOrd1}
\end{equation}
The amplitudes $A_i$ and $B_i$ for $i = 1,2$ are arbitrary (to this end) complex constants, while the phases $\psi$ and $\phi_{1,2}$ are given by 
\begin{equation}
\psi = \Omega \tau - K s, \qquad \qquad \phi_{1,2} = \pm \omega \tau - k s. \label{Phase}
\end{equation}
The dispersion relations between the wave numbers $k$ and $K$, and the wave frequencies $\omega$ and $\Omega$ can be expressed as 
\begin{equation}
\hbar^2 {\left( \Omega^2 - K^2 \right)} = 1 + A_0^2, \qquad \qquad \omega^2 = {\left( k - k_0 \right)}^2. \label{DispRel}
\end{equation}
The fact that the constants $B_{1,2}$ are strictly speaking complex becomes clearly visible from a more advantageous transform of the Klein-Gordon equation (which is of second order in the time variable) into a system of two coupled differential equations that are of first order in time. More details concerning this so-called Feshbach-Villars \cite{Feshbach} transformation can be found in Appendix \ref{sec:appendixA}. 

The aim of the present article is to demonstrate the nonlinear coupling processes between the quantum electron beam modes and the generated electromagnetic radiation as a result of the passage of the electron beam through the undulator. For simplicity and comprehensibility of the subsequent exposition we shall consider the following single-mode case
\begin{equation}
{\cal A}_1 = A e^{i \phi}, \qquad \phi = \phi_1, \qquad \qquad \Psi_1 = B e^{i \psi}. \label{SolEquFirOrd}
\end{equation}
In addition, let us relate the energy and the momentum of the electron beam to the corresponding quantum mechanical values  
\begin{equation}
\Gamma_0 = \hbar \Omega, \qquad \qquad \Gamma_0 v_0 = \hbar K. \label{StatInit}
\end{equation}
where $\Gamma_0$ and $\Gamma_0 v_0$ are the unperturbed values of the energy and the momentum, respectively. Note that according to Eqs. (\ref{Gamma}) and (\ref{DispRel}), we must have 
\begin{equation}
\Gamma_0^2 {\left( 1 - v_0^2 \right)} = 1 + A_0^2. \label{EnergyCon}
\end{equation}
It should be mentioned that once the first-order electron wave function $\Psi_1$ is available through the second expression in Eq. (\ref{SolEquFirOrd}), we can determine the stationary value of the electron current velocity. Since  
\begin{equation}
n_2 = \Gamma_0 {\left| B \right|}^2, \qquad \qquad j_{s2} = \Gamma_0 v_0 {\left| B \right|}^2, \label{SecOrdHydro}
\end{equation}
we identify the initial hydrodynamic current velocity $v_{s0}$ with the electron beam velocity $v_0$ and the enthalpy variable $\Gamma_0$ with the total beam-undulator energy. This fact is to be used in what follows. 

\subsubsection{\label{subsubsec:secondorder}Second Order} 

In explicit form the second-order equations (\ref{WaveEqu2}) and (\ref{KleinGordon}) can be written as follows 
\begin{equation}
{\boldsymbol{\Box}} {\cal A}_2 + 2 i k_0 \partial_s {\cal A}_2 - k_0^2 {\cal A}_2 = {\frac {\mu_0 e^2 A_0} {m_e \Gamma_0}} n_2, \label{WaveEquSecOrd}
\end{equation}
\begin{equation}
\hbar^2 {\boldsymbol{\Box}} \Psi_2 - {\left( 1 + A_0^2 \right)} \Psi_2 = A_0 {\left( {\cal A}_1 + {\cal A}_1^{\ast} \right)} \Psi_1. \label{KleinGordSecOrd}
\end{equation}
Moreover, the second-order generalized potential $V_2$ satisfies the equation 
\begin{equation}
\partial_s^2 V_2 = {\frac {\mu_0 e^2} {m_e}} n_2, \label{LaplaceEqSO}
\end{equation}
where the second-order electron line density $n_2$ is given by the first of Eqs. (\ref{SecOrdHydro}). The solutions to the second-order equations displayed above can be written in the form 
\begin{equation}
{\cal A}_2 = - {\frac {\mu_0 e^2 A_0} {m_e k_0^2}} {\left| B \right|}^2, \label{SolEquSecOrd1}
\end{equation}
\begin{equation}
\Psi_2 = A_0 {\left( \Lambda_1 {\cal A}_1 + \Lambda_2 {\cal A}_1^{\ast} \right)} \Psi_1, \label{SolEquSecOrd2}
\end{equation}
where 
\begin{eqnarray}
\Lambda_1 = {\frac {1} {\hbar^2 k_0^2 - 2 \hbar^2 k_0 k + 2 \hbar \omega \Gamma_0 - 2 \hbar k \Gamma_0 v_0}}, \label{SolLambda1} 
\\ 
\Lambda_2 = {\frac {1} {\hbar^2 k_0^2 - 2 \hbar^2 k_0 k - 2 \hbar \omega \Gamma_0 + 2 \hbar k \Gamma_0 v_0}}. \label{SolLambda2}
\end{eqnarray}

\subsubsection{\label{subsubsec:thirdorder}Third Order. Derivation of the Nonlinear Amplitude Equation} 

The explicit form of the third-order equations (\ref{WaveEqu2}) and (\ref{KleinGordon}) can be represented as follows 
\begin{equation}
{\boldsymbol{\Box}} {\cal A}_3 + 2 i k_0 \partial_s {\cal A}_3 - k_0^2 {\cal A}_3 = {\frac {\mu_0 e^2} {m_e \Gamma_0}} {\left( n_2 {\cal A}_1 + n_3 A_0 \right)}, \label{WaveEquThiOrd}
\end{equation}
\begin{eqnarray}
\hbar^2 {\boldsymbol{\Box}} \Psi_3 - {\left( 1 + A_0^2 \right)} \Psi_3 = -2i \hbar V_2 \partial_{\tau} \Psi_1 - i \hbar \Psi_1 \partial_{\tau} V_2 \nonumber
\\ 
+ A_0 {\left( {\cal A}_1 + {\cal A}_1^{\ast} \right)} \Psi_2 + {\left( {\left| {\cal A}_1 \right|}^2 + 2 A_0 {\cal A}_2 \right)} \Psi_1. \label{KleinGordThiOrd}
\end{eqnarray}
Here the third-order electron beam number density reads as 
\begin{equation}
n_3 = {\frac {i \hbar} {2}} {\left( \Psi_1 \partial_{\tau} \Psi_2^{\ast} + \Psi_2 \partial_{\tau} \Psi_1^{\ast} - \Psi_1^{\ast} \partial_{\tau} \Psi_2 - \Psi_2^{\ast} \partial_{\tau} \Psi_1 \right)}. \label{ThiOrdNumDen}
\end{equation}

Unlike the second-order case considered above, where the perturbation equations possess unique regular solutions, there are two types of terms on the right-hand-side of the third-order equations (\ref{WaveEquThiOrd}) and (\ref{KleinGordThiOrd}), as can be easily observed. The first type comprises a collection of secular ({\it aka} resonant) terms, which follow the pattern of the basic wave mode(s) in linear approximation. Such terms, which appear in higher orders as well, would provide a divergent counterpart in the naive perturbation solution. Thus, they must be renormalized by an elegant renormalization group procedure described below. The rest of the terms contribute to the regular solution of the third-order perturbation equations, involving higher harmonics and/or higher order harmonic linear combinations/superpositions of the modes proportional to $e^{\pm i \phi}$ and $e^{\pm i \psi}$. Omitting straightforwardly reproducible calculation’s details, we write down the resonant part of equations (\ref{WaveEquThiOrd}) and (\ref{KleinGordThiOrd}) 
\begin{eqnarray}
{\boldsymbol{\Box}} {\cal A}_3 + 2 i k_0 \partial_s {\cal A}_3 - k_0^2 {\cal A}_3 = {\frac {\mu_0 e^2} {m_e \Gamma_0}} n_2 {\cal A}_1 \nonumber 
\\ 
+ {\frac {\mu_0 e^2 A_0^2} {2 m_e \Gamma_0}} 
{\left[ \Lambda_1 {\left( 2 \Gamma_0 + \hbar \omega \right)} + \Lambda_2 {\left( 2 \Gamma_0 - \hbar \omega \right)} \right]} {\left| B \right|}^2 {\cal A}_1, \label{WaEqThiOrdSec}
\end{eqnarray}
\begin{eqnarray}
\hbar^2 {\boldsymbol{\Box}} \Psi_3 - {\left( 1 + A_0^2 \right)} \Psi_3 = 2 \Gamma_0 V_2 \Psi_1 - i \hbar \Psi_1 \partial_{\tau} V_2 \nonumber
\\ 
+ 2 A_0^2 \hbar^2 \Lambda_1 \Lambda_2 {\left( \omega^2 - k^2 \right)} {\left| A \right|}^2 \Psi_1 + {\left( {\left| A \right|}^2 + 2 A_0 {\cal A}_2 \right)} \Psi_1. \label{KlGordThiOrdSec}
\end{eqnarray}
The above equations possess an exact solution, which can be expressed as 
\begin{equation}
{\cal A}_3 {\left( s; \tau \right)} = P_3 {\left( s; \tau \right)} e^{i \phi}, \qquad \Psi_3 {\left( s; \tau \right)} = Q_3 {\left( s; \tau \right)} e^{i \psi}, \label{ProtoRGSol}
\end{equation}
where the amplitudes $P_3 {\left( s; \tau \right)}$ and $Q_3 {\left( s; \tau \right)}$ satisfy the equations
\begin{eqnarray}
{\boldsymbol{\Box}} P_3 - 2 i {\left[ \omega \partial_{\tau} + {\left( k - k_0 \right)} \partial_s \right]} P_3 = {\frac {\mu_0 e^2} {m_e \Gamma_0}} n_2 A \nonumber 
\\ 
+ {\frac {\mu_0 e^2 A_0^2} {2 m_e \Gamma_0}} 
{\left[ \Lambda_1 {\left( 2 \Gamma_0 + \hbar \omega \right)} + \Lambda_2 {\left( 2 \Gamma_0 - \hbar \omega \right)} \right]} {\left| B \right|}^2 A, \label{WaEqThiOrdRen}
\end{eqnarray}
\begin{eqnarray}
\hbar^2 {\boldsymbol{\Box}} Q_3 - 2 i \hbar \Gamma_0 {\left( \partial_{\tau} + v_0 \partial_s \right)} Q_3 = 2 \Gamma_0 V_2 B - i \hbar B \partial_{\tau} V_2 \nonumber
\\ 
+ 2 A_0^2 \hbar^2 \Lambda_1 \Lambda_2 {\left( \omega^2 - k^2 \right)} {\left| A \right|}^2 B + {\left( {\left| A \right|}^2 + 2 A_0 {\cal A}_2 \right)} B. \label{KlGordThiOrdRen}
\end{eqnarray}

We follow an elegant approach, known as the proto RG operator scheme \cite{Oono,Nozaki,Shiwa}, which has been proposed in the early 2000s to free as much as possible the standard RG theoretical reduction from the necessity of explicit (in the majority of cases, rather cumbersome) calculation of secular terms. As a result, we finally obtain the following amplitude equations 
\begin{eqnarray}
{\boldsymbol{\Box}} A - 2 i {\left[ \omega \partial_{\tau} + {\left( k - k_0 \right)} \partial_s \right]} A = {\frac {\mu_0 e^2} {m_e \Gamma_0}} n_2 A \nonumber 
\\ 
+ {\frac {\mu_0 e^2 A_0^2} {2 m_e \Gamma_0}} 
{\left[ \Lambda_1 {\left( 2 \Gamma_0 + \hbar \omega \right)} + \Lambda_2 {\left( 2 \Gamma_0 - \hbar \omega \right)} \right]} {\left| B \right|}^2 A, \label{WaEqAmplitude}
\end{eqnarray}
\begin{eqnarray}
\hbar^2 {\boldsymbol{\Box}} B - 2 i \hbar \Gamma_0 {\left( \partial_{\tau} + v_0 \partial_s \right)} B = 2 \Gamma_0 V_2 B - i \hbar B \partial_{\tau} V_2 \nonumber
\\ 
+ 2 A_0^2 \hbar^2 \Lambda_1 \Lambda_2 {\left( \omega^2 - k^2 \right)} {\left| A \right|}^2 B + {\left( {\left| A \right|}^2 + 2 A_0 {\cal A}_2 \right)} B. \label{KlGordAmplitude}
\end{eqnarray}

Equations (\ref{WaEqAmplitude}) and (\ref{KlGordAmplitude}) comprise a system of coupled nonlinear wave equations governing the evolution of the slowly varying amplitudes of the electron beam quantum density distribution and the complex transverse component of the electromagnetic vector potential describing the radiation field. 

\section{\label{sec:linearstab}Linear Stability}

Following Takabayasi \cite{Takabayasi}, it is convenient to introduce a hydrodynamic formulation of the Klein-Gordon equation (\ref{KleinGordon}) in terms of the eikonal decomposition 
\begin{equation}
\Psi {\left( s; \tau \right)} = R {\left( s; \tau \right)} e^{i S {\left( s; \tau \right)} / \hbar}, \label{Eikonal}
\end{equation}
where the amplitude $R {\left( s; \tau \right)}$ and the phase $S {\left( s; \tau \right)}$ are real functions of the arguments indicated. Separating the real and imaginary parts of the Klein-Gordon equation, we have
\begin{equation}
{\left( \partial_{\tau} S - V \right)}^2 - {\left( \partial_s S \right)}^2 - 1 - {\left| {\cal A} \right|}^2 = - \hbar^2 {\frac {{\boldsymbol{\Box}} R} {R}}, \label{Madelung1}
\end{equation}
\begin{equation}
R {\left( {\boldsymbol{\Box}} S + \partial_{\tau} V \right)} - 2 {\left( \partial_{\tau} R \right)} {\left( \partial_{\tau} S - V \right)} + 2 {\left( \partial_s R \right)} {\left( \partial_s S \right)}  = 0. \label{Madelung2}
\end{equation}
In terms of the amplitude $R {\left( s; \tau \right)}$ and the phase $S {\left( s; \tau \right)}$, the number $n$ and current $j_s$ densities are, respectively 
\begin{equation}
n = R^2 {\left( \partial_{\tau} S - V \right)}, \qquad \quad j_s = n v_s = - R^2 \partial_s S. \label{NumDens}
\end{equation}

Notwithstanding that the eikonal substitution (\ref{Eikonal}) is similar to that introduced by Madelung \cite{Madelung} for the Schrodinger equation, the form of the quantum hydrodynamic equations is quite different, but still, quite convenient for a more detailed further analysis. Alternative fluid dynamic methods for the Klein-Gordon equation are also available, such as the Feshbach-Villars \cite{Feshbach} formalism briefly discussed in Appendix \ref{sec:appendixA}. The resulting set of hydrodynamic equations turns out to appear much more complicated than in the classic Takabayasi approach, which we shall use here due to its formal simplicity. 

Here is the place to make an important lyrical digression. Let us introduce the particle density $\varrho$ and covariant flow velocity $u_{\mu}$ according to the expressions 
\begin{equation}
\varrho = R^2, \qquad \quad u_{\mu} = {\left( \partial_{\tau} S - V, a_x, a_y, - \partial_s S \right)}. \label{Covariant}
\end{equation}
In this way, we can rewrite Eqs. (\ref{Madelung1}) and (\ref{Madelung2}) in a covariant form as follows 
\begin{equation}
\partial_{\mu} {\left( \varrho u^{\mu} \right)} = 0, \qquad \qquad u_{\mu} u^{\mu} - 1 = - \hbar^2 {\frac {{\boldsymbol{\Box}} {\sqrt{\varrho}}} {{\sqrt{\varrho}}}}. \label{CovarMadelung}
\end{equation}
If we take the covariant derivative of the second of Eqs. (\ref{CovarMadelung}), the resulting equations become exactly those of the relativistic hydrodynamics of the charged fluid with the additional quantum force term. 

\subsubsection{\label{subsubsec:equilib}Equilibrium Solution} 

The system of equations (\ref{Madelung1}) -- (\ref{NumDens}) together with Eq. (\ref{WaveEqu2}) possess an equilibrium solution 
\begin{equation}
S_0 = \Gamma_0 \tau - \Gamma_0 v_0 s, \qquad \qquad R_0 = {\sqrt{n_0}}, \label{Equilibr1}
\end{equation}
\begin{equation}
{\cal A}_0 = A_0 = - {\frac {\omega_c k_0 c} {\omega_p^2 + k_0^2 c^2}}, \label{Equilibr2}
\end{equation}
where 
\begin{equation}
\omega_p^2 = {\frac {e^2 n_0} {\epsilon_0 m_e}}, \label{PlasmaFreq}
\end{equation}
is the plasma frequency. Moreover, $n_0$ is the equilibrium beam number density, and as expected the relation (\ref{EnergyCon}) is automatically recovered by Eq. (\ref{Madelung1}). Also note that in obtaining the stationary solutions above, the explicit assumption that the electron beam density $n_0$ is relatively small has been made. Therefore, the longitudinal space-charge effects can be neglected, that is $V_0 = 0$.

\subsubsection{\label{subsubsec:linearized}Linearized Equations and the High-Gain FEL Regime} 

It is convenient to split the enthalpy variable $\Gamma_0$ into two parts according to 
\begin{equation}
\Gamma_0 = \gamma_0 {\mathfrak R}_0, \qquad \gamma_0 = {\frac {1} {\sqrt{1 - v_0^2}}}, \qquad {\mathfrak R}_0 = {\sqrt{1 + A_0^2}}. \label{Enthalpy}
\end{equation}
By linearizing the hydrodynamic equations (\ref{Madelung1}) -- (\ref{NumDens}) together with the field equations (\ref{WaveEqu2}) and (\ref{LaplaceEq}), we obtain 
\begin{eqnarray}
{\frac {\hbar^2} {R_0}} {\boldsymbol{\Box}} R_1 + 2 \gamma_0 {\mathfrak R}_0 {\left( \partial_{\tau} + v_0 \partial_s \right)} S_1 - 2 \gamma_0 {\mathfrak R}_0 V_1 \nonumber 
\\ 
= A_0 {\left( {\cal A}_1 + {\cal A}_1^{\ast} \right)} \label{MadelLin1}
\end{eqnarray}
\begin{equation}
R_0 {\boldsymbol{\Box}} S_1 - 2 \gamma_0 {\mathfrak R}_0 {\left( \partial_{\tau} + v_0 \partial_s \right)} R_1 + R_0 \partial_{\tau} V_1 = 0. \label{MadeLin2}
\end{equation}
\begin{equation}
n_1 = n_0 {\left( \partial_{\tau} S_1 - V_1 \right)} + 2 \gamma_0 {\mathfrak R}_0 R_0 R_1, \label{NumDensLin}
\end{equation}
\begin{equation}
{\boldsymbol{\Box}} {\cal A}_1 + 2 i k_0 \partial_s {\cal A}_1 - {\left( k_0^2 + {\frac {\omega_p^2} {c^2}} \right)} {\cal A}_1 = {\frac {A_0 \omega_p^2} {\Gamma_0 c^2}} {\frac {n_1} {n_0}}, \label{WaveEquLin2}
\end{equation}
\begin{equation}
\partial_s^2 V_1 = {\frac {\omega_p^2} {c^2}} {\frac {n_1} {n_0}}. \label{LaplaceEqLin}
\end{equation}

Eliminating the first-order eikonal variable $S_1$ by apparent application of the d'Alembert operator ${\boldsymbol{\Box}}$ on both sides of Eqs. (\ref{MadelLin1}) and (\ref{NumDensLin}) with due account of Eq. (\ref{MadeLin2}), we obtain 
\begin{eqnarray}
{\left( \hbar^2 {\boldsymbol{\Box}}^2 + 4 \gamma_0^2 {\mathfrak R}_0^2 {\widehat{\pmb{\mathfrak{D}}}}^2 \right)} R_1 \nonumber 
\\ 
= 2 \gamma_0 {\mathfrak R}_0 R_0 {\widehat{\pmb{\mathfrak{D}}}}_1 \partial_s V_1 + A_0 R_0 {\boldsymbol{\Box}} {\left( {\cal A}_1 + {\cal A}_1^{\ast} \right)}, \label{CoupLin1}
\end{eqnarray}
\begin{equation}
{\left( {\boldsymbol{\Box}} + {\frac {\omega_p^2} {c^2}} \right)} n_1 = 2 \gamma_0 {\mathfrak R}_0 R_0 {\widehat{\pmb{\mathfrak{D}}}}_1 \partial_s R_1, \label{CoupLin2}
\end{equation}
respectively. Here we have introduced the operators 
\begin{equation}
{\widehat{\pmb{\mathfrak{D}}}} = \partial_{\tau} + v_0 \partial_s, \qquad \qquad {\widehat{\pmb{\mathfrak{D}}}}_1 = v_0 \partial_{\tau} + \partial_s. \label{OperatorsD}
\end{equation}
The first-order amplitude $R_1$ (and in addition, the first-order space-charge potential $V_1$, as is involved in the process of the detailed algebraic manipulations) can be easily eliminated from the Eqs. (\ref{CoupLin1}) and (\ref{CoupLin2}) to finally arrive at the desired result 
\begin{eqnarray}
{\boldsymbol{\Box}} {\left[ \hbar^2 {\boldsymbol{\Box}} {\left( {\boldsymbol{\Box}} + {\frac {\omega_p^2} {c^2}} \right)} + 4 {\mathfrak R}_0^2 {\left( \gamma_0^2 {\widehat{\pmb{\mathfrak{D}}}}^2 - {\frac {\omega_p^2} {c^2}} \right)} \right]} n_1 \nonumber 
\\ 
= 2 \gamma_0 {\mathfrak R}_0 n_0 A_0 {\boldsymbol{\Box}} {\widehat{\pmb{\mathfrak{D}}}}_1 \partial_s {\left( {\cal A}_1 + {\cal A}_1^{\ast} \right)}. \label{FinalCoup}
\end{eqnarray}
This last equation is coupled to Eq. (\ref{WaveEquLin2}), which can be written in a form 
\begin{eqnarray}
{\left[ {\left( {\boldsymbol{\Box}} - k_1^2 \right)}^2 + 4 k_0^2 \partial_s^2 \right]} {\left( {\cal A}_1 + {\cal A}_1^{\ast} \right)} \nonumber 
\\ 
= {\frac {2 A_0 \omega_p^2} {\Gamma_0 n_0 c^2}} {\left( {\boldsymbol{\Box}} - k_1^2 \right)} n_1, \label{WaveEquLinFin}
\end{eqnarray}
containing real quantities and variables only. Here, the notation  
\begin{equation}
k_1^2 = k_0^2 + {\frac {\omega_p^2} {c^2}}, \label{Constk1}
\end{equation}
has been introduced. It is now a simple matter to obtain a single equation for the horizontal component ${\cal A}_1 + {\cal A}_1^{\ast}$ of the vector potential. The result is
\begin{equation}
{\widehat{\pmb{\mathfrak{N}}}} {\boldsymbol{\Box}} {\left( {\cal A}_1 + {\cal A}_1^{\ast} \right)} = 0, \label{DispersEquat}
\end{equation}
where 
\begin{eqnarray}
{\widehat{\pmb{\mathfrak{N}}}} = {\left[ \hbar^2 {\boldsymbol{\Box}} {\left( {\boldsymbol{\Box}} + {\frac {\omega_p^2} {c^2}} \right)} + 4 {\mathfrak R}_0^2 {\left( \gamma_0^2 {\widehat{\pmb{\mathfrak{D}}}}^2 - {\frac {\omega_p^2} {c^2}} \right)} \right]} \nonumber 
\\ 
{\left[ {\left( {\boldsymbol{\Box}} - k_1^2 \right)}^2 + 4 k_0^2 \partial_s^2 \right]} - {\frac {4 A_0^2 \omega_p^2} {c^2}} {\left( {\boldsymbol{\Box}} - k_1^2 \right)} {\widehat{\pmb{\mathfrak{D}}}}_1 \partial_s. \label{DispersOper}
\end{eqnarray}
Note that, the left-hand-side of Eq. (\ref{DispersEquat}) represents a product of two operators. The action of the d'Alembert operator ${\boldsymbol{\Box}}$ implies that a possible linear solution ${\cal A}_1 + {\cal A}_1^{\ast}$ to our fluid dynamic model just follows a harmonic wave pattern that simply satisfies the homogeneous wave equation. This fact is not surprising, since the hydrodynamic representation of the Klein-Gordon equation is invariant under Lorentz transformation. 

\subsubsection{\label{subsubsec:lindisper}Dispersion Equation and the Condition for Lasing} 

Next, we seek a solution to the linearized equation ${\widehat{\pmb{\mathfrak{N}}}} {\left( {\cal A}_1 + {\cal A}_1^{\ast} \right)} = 0$ in terms of the usual linear superposition form
\begin{equation}
{\cal A}_1 + {\cal A}_1^{\ast} = \sum_n \int \limits_{-\infty}^{\infty} {\rm d} K {\mathfrak A} {\left( K \right)} \exp {\left\{ i {\left[ K s - \Omega_n {\left( K \right)} \right]} \tau \right\}}, \label{LinSuperpos}
\end{equation}
where $\Omega_n {\left( K \right)}$ are all possible solutions of the dispersion equation 
\begin{widetext}
\begin{eqnarray}
{\left\{ \hbar^2 {\left( \Omega^2 - K^2 \right)} {\left( \Omega^2 - K^2 + {\frac {\omega_p^2} {c^2}} \right)} - 4 {\mathfrak R}_0^2 {\left[ \gamma_0^2 {\left( \Omega - v_0 K \right)}^2 + {\frac {\omega_p^2} {c^2}} \right]} \right\}} {\left[ {\left( \Omega^2 - K^2 - k_1^2 \right)}^2 - 4 k_0^2 K^2 \right]} \nonumber
\\ 
+ {\frac {4 A_0^2 \omega_p^2} {c^2}} {\left( \Omega^2 - K^2 - k_1^2 \right)} {\left( K^2 - v_0 \Omega K \right)} = 0. \label{LinDispEquat}
\end{eqnarray}
\end{widetext}
Considering this equation as an equation for $\Omega$ as a function of $K$, we notice that it is a polynomial equation of the eighth degree, which, as is well known, is not solvable in quadratures. Furthermore, in the general case, the dispersion equation possesses eight distinct roots which can be distributed in one or more complex-conjugate pairs. The case of free electron laser instability, sometimes also called the lazing condition, is realized when at least one complex-conjugate pair of roots of the dispersion equation is present. In the present article, we will be interested in this case.

In order to assess the predictions of the quantum-hydrodynamics model developed here on a qualitative and quantitative level, the dispersion equation (\ref{LinDispEquat}) should be solved numerically. Note that in the notation adopted here, the wave frequency ${\left[ \Omega \right]} = m^{-1}$ is, in fact, the radiation wave number.

\begin{figure}
\begin{center} 
\includegraphics[width=8.0cm]{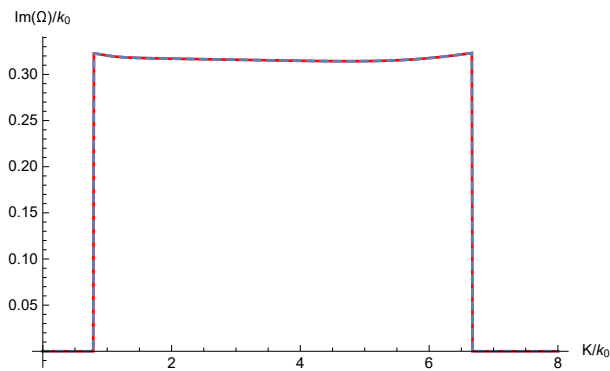}
\caption{\label{fig1:epsart} Normalized free electron laser instability growth rate ${\rm Im} {\left( \Omega \right)} / k_0$ as a function of the normalized radiation wave number $K / k_0$. The red curve represents the solution of the dispersion equation (\ref{LinDispEquat}), while the dashed blue curve represents its solution in quasi-classical approximation $\hbar \rightarrow 0$. The following particular values of the main parameters have been used: $\gamma_0 = 2$, $\lambda_0 = 1$ cm, $\omega_c/ {\left( k_0 c \right)} = 0.467$, ${\left[ \omega_p/ {\left( k_0 c \right)} \right]}^2 = 0.404$}
\end{center}
\end{figure}

\begin{figure}
\begin{center} 
\includegraphics[width=8.0cm]{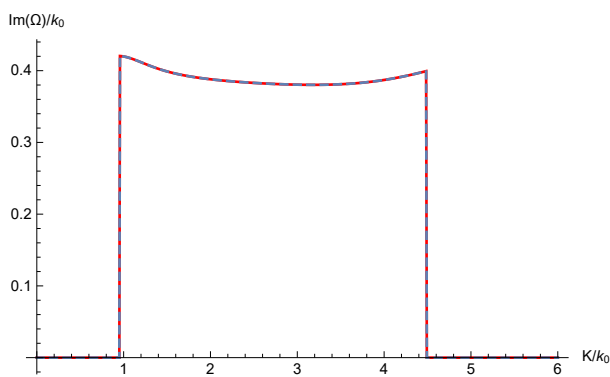}
\caption{\label{fig2:epsart} The same as in Figure \ref{fig1:epsart}, but at higher undulator magnetic field strength $B_0$ and electron beam density $n_0$ values. The following particular values of the main parameters have been used: $\gamma_0 = 2$, $\lambda_0 = 1$ cm, $\omega_c/ {\left( k_0 c \right)} = 2.334$, ${\left[ \omega_p/ {\left( k_0 c \right)} \right]}^2 = 0.583$}
\end{center}
\end{figure}

At low $\gamma_0$ values (Figures \ref{fig1:epsart} and \ref{fig2:epsart}), the observed behavior of the normalized free electron laser instability growth rate ${\rm Im} {\left( \Omega \right)} / k_0$ is qualitatively similar to that reported earlier \cite{TzenovMarinov}, where classical hydrodynamic approach involving exact hydrodynamic closure has been proposed. Here is the place, however, to note that quantum corrections practically do not affect the basic stroke and behavior of the instability bandwidth curve, as compared to those predicted in the quasi-classical approximation. 

At higher electron beam energy $\gamma_0$ and undulator magnetic field strength $B_0$ values (see Figure 3) the instability bandwidth increases and the location of its peak shifts towards shorter radiation wavelengths. Note that the match of the quantum with the quasi-classical description is still remarkable.
\begin{figure}
\begin{center} 
\includegraphics[width=8.0cm]{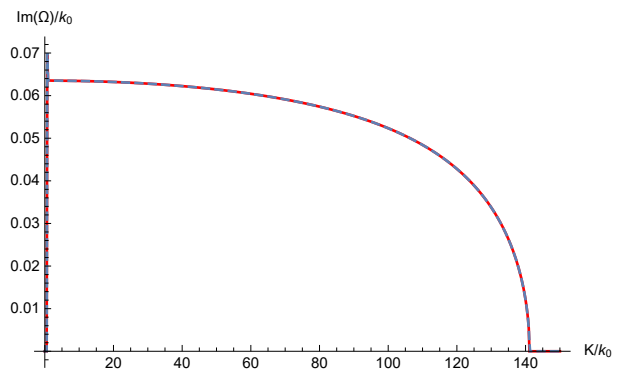}
\caption{\label{fig3:epsart} The same as in Figure \ref{fig1:epsart}, but at higher beam energy $\gamma_0$ and magnetic field strength $B_0$ values. The following particular values of the main parameters have been used: $\gamma_0 = 10$, $\lambda_0 = 1$ cm, $\omega_c/ {\left( k_0 c \right)} = 7$, ${\left[ \omega_p/ {\left( k_0 c \right)} \right]}^2 = 0.404$}
\end{center}
\end{figure}

\begin{figure}
\begin{center} 
\includegraphics[width=8.0cm]{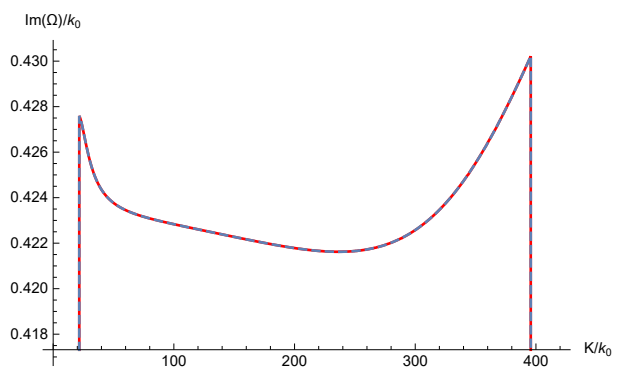}
\caption{\label{fig4:epsart} Dependence of the normalized instability growth rate ${\rm Im} {\left( \Omega \right)} / k_0$ on beam energy $\gamma_0$ for fixed magnetic field $B_0$, beam density $n_0$, and undulator period $\lambda_0$ values. The following particular values of the main parameters have been used: $\gamma_0 = 15$, $\lambda_0 = 10$ cm, $\omega_c/ {\left( k_0 c \right)} = 14$, ${\left[ \omega_p/ {\left( k_0 c \right)} \right]}^2 = 40$}
\end{center}
\end{figure}

\begin{figure}
\begin{center} 
\includegraphics[width=8.0cm]{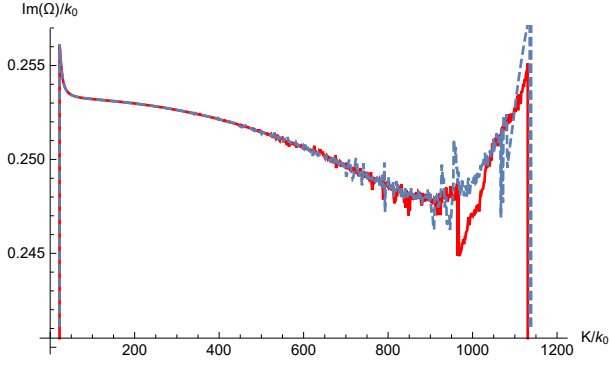}
\caption{\label{fig5:epsart} The same as in Figure \ref{fig4:epsart}, but at higher values of the beam energy $\gamma_0$. The following particular values of the main parameters have been used: $\gamma_0 = 25$, $\lambda_0 = 10$ cm, $\omega_c/ {\left( k_0 c \right)} = 14$, ${\left[ \omega_p/ {\left( k_0 c \right)} \right]}^2 = 40$}
\end{center}
\end{figure}

\begin{figure}
\begin{center} 
\includegraphics[width=8.0cm]{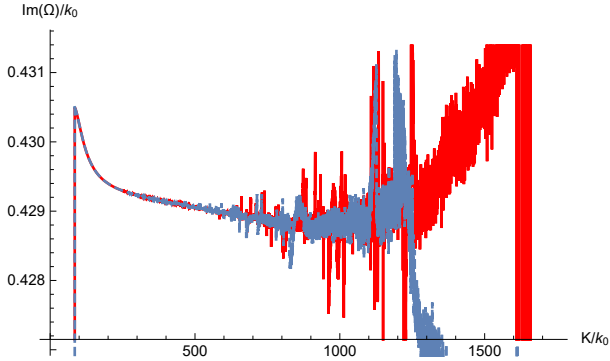}
\caption{\label{fig6:epsart} The same as in Figure \ref{fig5:epsart}, but the undulator period length has been increased by a factor of 2. The following particular values of the main parameters have been used: $\gamma_0 = 29.5$, $\lambda_0 = 20$ cm, $\omega_c/ {\left( k_0 c \right)} = 28$, ${\left[ \omega_p/ {\left( k_0 c \right)} \right]}^2 = 160$}
\end{center}
\end{figure}

Figures \ref{fig4:epsart}, \ref{fig5:epsart} and \ref{fig6:epsart} show the effect of beam energy $\gamma_0$ on the lasing condition at two different undulator period lengths $\lambda_0 = 10$ cm and $\lambda_0 = 20$ cm. As can be seen, at lower beam energies a single instability band with a couple of well-defined peaks exists, whereas at higher beam energies two (or more) separate bands are generated. The latter is due to the existence of more than one couple of complex-conjugate roots of the dispersion equation. 
Interestingly enough (see Figure \ref{fig6:epsart}), a threshold beam energy below which no instability exists, can be observed. By increasing the beam energy above its threshold value, an increase of both the bandwidth (range of $K$-values) and the peak value of $\Omega$ as a function of $K$ can be clearly identified. 

Another interesting and important feature of the obtained numerical results, which is immediately noticeable, is that at relatively low electron beam energies and intensities, the instability bandwidths calculated in the quasi-classical approximation are close enough to the quantum hydrodynamics ones. 

\section{\label{sec:concluding}Concluding Remarks} 

We present a relativistic quantum mechanical model to describe the quantum FEL dynamics. Neglecting the spin of electrons in the impacting beam, this model is based on the Klein-Gordon equation coupled to the Poisson equation for the space-charge potential and the wave equation for the transverse components of the radiation field. Using the elegant renormalization group method, a system of coupled nonlinear envelope equations for the slowly varying amplitudes of the electron beam distribution in the configuration space and the radiation field has been derived. 

The fundamental system of basic equations consisting of the Klein-Gordon equation for the electron beam wave function and the equations for the space-charge and radiation fields have been cast into a suitable hydrodynamic formulation. In the framework of the hydrodynamic representation, a new dispersion relation has been derived and analyzed in both the quantum and the quasi-classical regimes, where the space-charge oscillation of the electron beam is taken into account. 

It has been shown that some basic features of the high-gain FEL regime at moderate electron beam energies, beam densities and undulator fields can be derived from the fluid dynamic description in the quasi-classical approximation. At higher values, however, the growing discrepancy between quantum and quasi-classical descriptions begins to noticeably stand out.

\begin{acknowledgments}
We wish to thank Prof. Zhentang Zhao for his continuous interest and support of the present article. Interesting and useful discussions with Drs. Bart Faatz and Jianhui Chen are also gratefully acknowledged. 
\end{acknowledgments}

\appendix

\section{\label{sec:appendixA}Feshbach-Villars Transformation of the Klein-Gordon Equation}

To cast the Klein-Gordon equation into the form of two coupled Schrodinger equations, it is necessary to resolve the wave function $\Psi$ into the components representing the two degrees of freedom implied by the Klein-Gordon equation (\ref{KleinGord}). The sought-for new wave function ${\widehat{\boldsymbol{\Psi}}}$ is then a uni-columnar matrix (or column vector) formed from these two components. 

For the sake of convenience, let us rewrite Eq. (\ref{KleinGord}) as  
\begin{equation}
{\widehat{\pmb{\mathscr{D}}}}^2 \Psi - {\widehat{\bf P}}^2 \Psi - \Psi = 0, \label{KleinGordOper}
\end{equation}
where 
\begin{equation}
{\widehat{\pmb{\mathscr{D}}}} = i \hbar \partial_{\tau} + V, \qquad \qquad {\widehat{\bf P}}^2 = {\left( - i \hbar \partial_s \right)}^2 + {\left| {\cal A} \right|}^2. \label{Operators}
\end{equation}
The two components mentioned above can be introduced according to the expressions 
\begin{equation}
\Psi = \chi_1 + \chi_2, \qquad \qquad {\widehat{\pmb{\mathscr{D}}}} \Psi = \chi_1 - \chi_2. \label{ComponentsWF}
\end{equation}
It is easily proved that the two coupled differential equations 
\begin{equation}
{\widehat{\pmb{\mathscr{D}}}} \chi_1 - {\frac {1} {2}} {\widehat{\bf P}}^2 \Psi - \chi_1 = 0, \label{Schrod1}
\end{equation}
\begin{equation}
{\widehat{\pmb{\mathscr{D}}}} \chi_2 + {\frac {1} {2}} {\widehat{\bf P}}^2 \Psi + \chi_2 = 0, \label{Schrod2}
\end{equation}
are equivalent to the Klein-Gordon equation (\ref{KleinGordOper}).

The coupled equations (\ref{Schrod1}) and (\ref{Schrod2}) may be combined to form one single equation of Schrodinger type 
\begin{equation}
i \hbar \partial_{\tau} {\widehat{\boldsymbol{\Psi}}} = {\widehat{\pmb{\mathscr{H}}}} {\widehat{\boldsymbol{\Psi}}}. \label{SchrodEquat}
\end{equation}
Here, we have introduced the column vector
\begin{equation}
{\widehat{\boldsymbol{\Psi}}} = 
\begin{pmatrix}
\chi_1 \\
\chi_2 
\end{pmatrix}. \label{WFColVec}
\end{equation}
In addition, the matrix Hamiltonian ${\widehat{\pmb{\mathscr{H}}}}$ reads as 
\begin{equation}
{\widehat{\pmb{\mathscr{H}}}} = {\frac {1} {2}} {\left( {\widehat{\boldsymbol{\sigma}}}_3 + i {\widehat{\boldsymbol{\sigma}}}_2 \right)} {\widehat{\bf P}}^2 - V {\widehat{\bf I}} + {\widehat{\boldsymbol{\sigma}}}_3, \label{MatrixHamil}
\end{equation}
where ${\widehat{\bf I}}$ is the unit $2 \times 2$ matrix, while 
\begin{equation}
{\widehat{\boldsymbol{\sigma}}}_1 =
\begin{pmatrix}
0 & 1 \\
1 & 0
\end{pmatrix}, \qquad 
{\widehat{\boldsymbol{\sigma}}}_2 =
\begin{pmatrix}
0 & -i \\
i & 0
\end{pmatrix}, \qquad 
{\widehat{\boldsymbol{\sigma}}}_3 =
\begin{pmatrix}
1 & 0 \\
0 & -1
\end{pmatrix}, \label{PauliMatrix}
\end{equation}
are the well-known Pauli matrices with the significant difference that they do not act in spin space, rather than in the vector space defined by Eq. (\ref{WFColVec}). In the Feshbach-Villars representation the expression for the electron current density (\ref{Density}) becomes especially simple
\begin{equation}
n = {\left| \chi_2 \right|}^2 - {\left| \chi_1 \right|}^2 = - {\widehat{\boldsymbol{\Psi}}}^{\dag} {\widehat{\boldsymbol{\sigma}}}_3 {\widehat{\boldsymbol{\Psi}}}, \label{DensitFV}
\end{equation}
where the "$\dag$" denotes the Hermitian conjugate. 

Let us consider the "free electron" case described by Eq. (\ref{KleinGordFirOrd}) in the Feshbach-Villars representation. If we write 
\begin{equation}
{\widehat{\boldsymbol{\Psi}}} = B 
\begin{pmatrix}
\chi_{10} \\
\chi_{20} 
\end{pmatrix} e^{i {\left( \Omega \tau - k s \right)}}, \label{FreeElect}
\end{equation}
and substitute this expression into Eq. (\ref{SchrodEquat}), we obtain 
\begin{equation}
{\widehat{\pmb{\mathscr{M}}}}
\begin{pmatrix}
\chi_{10} \\
\chi_{20} 
\end{pmatrix} = 0, \label{FreeElSol}
\end{equation}
where the matrix ${\widehat{\pmb{\mathscr{M}}}}$ is expressed as 
\begin{equation} 
\begin{pmatrix}
\hbar \Omega + {\dfrac {1} {2}} {\left( \hbar^2 k^2 + A_0^2 \right)} + 1 & {\dfrac {1} {2}} {\left( \hbar^2 k^2 + A_0^2 \right)} \\
- {\dfrac {1} {2}} {\left( \hbar^2 k^2 + A_0^2 \right)} & \hbar \Omega - {\dfrac {1} {2}} {\left( \hbar^2 k^2 + A_0^2 \right)} - 1
\end{pmatrix}. \label{MatSol}
\end{equation}
From the condition that ${\rm det} {\widehat{\pmb{\mathscr{M}}}} = 0$ we recover the first of the dispersion relations (\ref{DispRel}). In addition $\chi_{10} = {\mathfrak C} {\left( 1 - \hbar \Omega \right)}$, and $\chi_{20} = {\mathfrak C} {\left( 1 + \hbar \Omega \right)}$, where ${\mathfrak C}$ is an arbitrary constant. If ${\mathfrak C} = 1 / 2$, the relations (\ref{SecOrdHydro}) are recovered as well.

It is also important to note that $\Omega$ can be both positive and negative, corresponding to positive and negative energy, respectively. It can be easily verified, the "negative" solution $\Omega \longrightarrow - {\left| \Omega \right|}$ is the solution charge conjugate to the "positive" solution $\Omega \longrightarrow {\left| \Omega \right|}$, so that if the positive solution represents a particle of positive charge, the $\Omega \longrightarrow - {\left| \Omega \right|}$ solution represents the one of negative charge. 






\bibliography{aipsamp}

\end{document}